\documentclass[aps,graphicx,12pt,showkeys,showpacs]{revtex4}
\usepackage{amsmath}
\usepackage{amscd}
\usepackage{graphicx}
\begin{document}

\title[Short Title]{One-step implementation of the Fredkin gate via quantum Zeno dynamics}

\author{Zhi-Cheng Shi$^{1}$}
\author{Yan Xia$^{1,2, }$\footnote{corresponding author E-mail: xia-208@163.com}}
\author{Jie Song$^{3}$}

\affiliation{$^{1}$Department of Physics, Fuzhou University, Fuzhou
350002, China \\ $^{2}$School of Physics and Optoelectronic
Technology, Dalian University of Technology, Dalian 116024,
China\\$^{3}$Department of Physics, Harbin Institute of Technology,
Harbin 150001, China}

\begin{abstract}We study one-step implementation of the Fredkin gate in a bi-modal cavity
under both resonant and large detuning conditions based on quantum
Zeno dynamics, which reduces the complexity of experiment
operations. The influence of cavity decay and atomic spontaneous
emission is discussed by numerical calculation. The results
demonstrate that the fidelity and the success probability are robust
against cavity decay in both models and they are also insensitive to
atomic spontaneous emission in the large detuning model. In
addition, the interaction time is rather short in the resonant model
compared to the large detuning model.
\end{abstract}

\pacs {03. 65. Xp, 03. 67. Lx}

 \maketitle

\section{INTRODUCTION}

Recently, attentions have been mainly paid to physical
implementation of quantum computer as it can provide a tremendous
speedup when it is compared to classical computer \cite{LK-PRL80},
such as searching for data in an array. Until now, much significant
progresses have been made in quantum computation during the last
decade. It is well known that the cavity quantum electrodynamics
(C-QED) which concerns the interaction of atoms and photons within
cavities is the promising candidate for quantum information
processing (QIP). There are numerous schemes proposed in the context
of C-QED for achieving quantum computation, such as a scheme for
realizing quantum logic gates and teleportation in C-QED
\cite{SB-PRL85}, a scheme for realizing two-qubit quantum phase gate
with a four-level system in C-QED \cite{CP-PRA70}, quantum logic
gates for two atoms with a single resonant interaction
\cite{SB-PRA71}, quantum phase gates for two atoms trapped in
separate cavities connected by an optical fiber \cite{ZB-PRA80},
etc..

For the case of three-qubit gates, much substantial efforts have
been made in the study of the fundamental Toffoli \cite{TM-PRL102}
and Fredkin \cite{GJ-PRL62} gates . The quantum Toffoli gate
performs a NOT operation on a target qubit depending on the states
of two control qubits while the quantum Fredkin gate performs a SWAP
operation on two target qubits depending on the state of a control
qubit. A number of schemes \cite{JF-PRA73,TC-PRA75,XQ-PRA75} were
proposed to implementation of a Toffoli gate and it was demonstrated
in the experiment successfully \cite{BP-arx0804}. The major
progresses made in the Toffoli gate provide a attractive motivation
for the attempts to implement the other three-qubit gate--Fredkin
gate. So far, there are numerous schemes
\cite{JF-PRA73,YX-PRA78,JF-PRA78} proposed to implementation of
Fredkin gate which is based on linear optical system. However, few
schemes \cite{XQS-PS79} are proposed to implementation of a Fredkin
gate based on atomic systems. Atomic systems are suitable to act as
qubits because moderate internal electronic states can coherently
store information over very long time scale. At the same time, the
switching on and off the atom-field interaction can be realizable
through simple controlling classical laser field. The remarkable
merits enlighten us to take a crack at constructing a theoretical
model for implementing Fredkin gate in C-QED.

On the other hand, quantum Zeno effect (QZE) is an fascinating
phenomenon which is applied to suppress decoherence via hindering
transition between quantum states by performing frequent
measurements \cite{09JMP18}. It can be showed in two mainly aspects.
On the one hand, the system will evolve away from its initial state
and remains in the Zeno subspace determined by the measurement when
frequently projected onto a multi-dimensional subspace
\cite{11PLA275,12PRA65}, which is called quantum Zeno dynamics. On
the other hand, Facchi \emph{et al.} \cite{13PIO42} showed that QZE
can also be reformulated in terms of a continuous coupling to obtain
the same purpose without making use of von Neumann's projections and
non-unitary dynamics in 2001. So far, numerous schemes have been
proposed to implementation of quantum computation
\cite{AB-PRL85,PJ-PRL89,KP-PRA69,XQ-PLA374,SZ-PBA44,ZCS-JOSAB28} via
QZE.

In this paper, we study one-step implementation of a Fredkin gate in
a bi-modal cavity under both resonant and large detuning conditions,
which is based on quantum Zeno dynamics. The advantages in both
models are threefold: (1) The gate can be implementation only one
step without any single-qubit gate operation, which reduces the
complexity of experiment operations. (2) The quantum information is
encoded in the low states. Thus there is no energy relaxation for
the atoms in the bi-modal cavity when the Fredkin gate operation is
finished. (3) The state keeps in a subspace without exciting the
cavity field during the whole system evolution, thus the fidelity
and success probability are robust against cavity decay. On the
other hand, in the large detuning model, the large detuning
condition eliminates the excited state of atoms adiabatically, which
is also insensitive to atomic spontaneous emission. The interaction
time is rather short in the resonant model compared to the large
detuning model.

This paper is organized as follows: In Sec. II, we study one-step
implementation of the Fredkin gate in a bi-modal cavity under
resonant conditions and large detuning based on quantum Zeno
dynamics, respectively. In Sec. III, we analyze the influence of
decoherence on the fidelity and success probability of the Fredkin
gate by numerical calculation. A discuss on experimental feasibility
and a summary are given in Sec. IV.

\section{implementation of a Fredkin gate in a bi-modal cavity based on quantum Zeno
dynamics}

\subsection{Under the resonant condition}

As shown in Fig. 1, three identical atoms, which have an excited
state $|e_{0}\rangle$, three ground states $|g_{L}\rangle$,
$|g_{R}\rangle$ and $|g_{0}\rangle$, interact with a bi-modal
cavity. The transitions $|e_{0}\rangle_k \leftrightarrow
|g_{L}\rangle_k (|g_{R}\rangle_k)$ (\emph{k}=1,2,3) are resonantly
coupled to left-circularly (right-circularly) polarized cavity
modes. The transition $|e_{0}\rangle_1\leftrightarrow
|g_{0}\rangle_1$ is resonantly driven by classical laser field with
Rabi frequency $\Omega$. In the interaction picture, the Hamiltonian
for the whole system can be written as ($\hbar = 1$)
\begin{eqnarray}\label{1}
H_{total}&=&H_{laser}+H_{c},\nonumber\\
 H_{laser}&=&\Omega(|e_{0}\rangle_{1}
\langle g_{0}|+|g_{0}\rangle_{1} \langle e_{0}|),\nonumber\\
H_{c}&=&\sum_{k=1}^{3}(g_{k,L}a_{L}|e_{0}\rangle_k\langle
g_{L}|+g_{k,R}a_{R}|e_{0}\rangle_k\langle g_{R}|)+H.c.
\end{eqnarray}
$a_{L}^\dagger$ ($a_{R}^\dagger$) and $a_{L}$ ($a_{R}$) are the
creation and annihilation operators for the bi-modal cavity mode.
$g_{k,L}$ ($g_{k,R}$) is the coupling strength between the
\emph{k}th atom and the left-(right-)circularly polarized cavity
mode. We assumed $g_{k,j}$ = $g\in R$ for simplicity. The quantum
information is encoded in a subspace spanned by the states
$\{|g_{R}\rangle_{1}$, $|g_{0}\rangle_{1}$, $|g_{L}\rangle_{2}$,
$|g_{R}\rangle_{2}$, $|g_{L}\rangle_{3}$, $|g_{R}\rangle_{3}\}$
while the cavity is in vacuum state $|00\rangle$.

For the initial states including
$|g_{R}\rangle_{1}|g_{L}\rangle_{2}|g_{L}\rangle_{3}|00\rangle$,
$|g_{R}\rangle_{1}|g_{L}\rangle_{2}|g_{R}\rangle_{3}|00\rangle$,
$|g_{R}\rangle_{1}|g_{R}\rangle_{2}|g_{L}\rangle_{3}|00\rangle$, and
$|g_{R}\rangle_{1}|g_{R}\rangle_{2}|g_{R}\rangle_{3}|00\rangle$,
they remain unchanged during the time evolution because
$H_{total}|g_{R}\rangle_{1}|g_{L}\rangle_{2}|g_{L}\rangle_{3}|00\rangle=0$,
$H_{total}|g_{R}\rangle_{1}|g_{L}\rangle_{2}|g_{R}\rangle_{3}|00\rangle=0$,
$H_{total}|g_{R}\rangle_{1}|g_{R}\rangle_{2}|g_{L}\rangle_{3}|00\rangle=0$,
and
$H_{total}|g_{R}\rangle_{1}|g_{R}\rangle_{2}|g_{R}\rangle_{3}|00\rangle=0$.

If the initial state is
$|g_{0}\rangle_{1}|g_{L}\rangle_{2}|g_{R}\rangle_{3}|00\rangle$, it
will evolve in a closed subspace spanned by
\begin{eqnarray}\label{2}
|\phi_{1}\rangle&=&|g_{0}\rangle_{1}|g_{L}\rangle_{2}|g_{R}\rangle_{3}|00\rangle,\nonumber\\
|\phi_{2}\rangle&=&|e_{0}\rangle_{1}|g_{L}\rangle_{2}|g_{R}\rangle_{3}|00\rangle,\nonumber\\
|\phi_{3}\rangle&=&|g_{L}\rangle_{1}|g_{L}\rangle_{2}|g_{R}\rangle_{3}|10\rangle,\nonumber\\
|\phi_{4}\rangle&=&|g_{L}\rangle_{1}|e_{0}\rangle_{2}|g_{R}\rangle_{3}|00\rangle,\nonumber\\
|\phi_{5}\rangle&=&|g_{L}\rangle_{1}|g_{R}\rangle_{2}|g_{R}\rangle_{3}|01\rangle,\nonumber\\
|\phi_{6}\rangle&=&|g_{L}\rangle_{1}|g_{R}\rangle_{2}|e_{0}\rangle_{3}|00\rangle,\nonumber\\
|\phi_{7}\rangle&=&|g_{L}\rangle_{1}|g_{R}\rangle_{2}|g_{L}\rangle_{3}|10\rangle,\nonumber\\
|\phi_{8}\rangle&=&|g_{R}\rangle_{1}|g_{L}\rangle_{2}|g_{R}\rangle_{3}|01\rangle,\nonumber\\
|\phi_{9}\rangle&=&|g_{R}\rangle_{1}|g_{L}\rangle_{2}|e_{0}\rangle_{3}|00\rangle,\nonumber\\
|\phi_{10}\rangle&=&|g_{R}\rangle_{1}|g_{L}\rangle_{2}|g_{L}\rangle_{3}|10\rangle,\nonumber\\
|\phi_{11}\rangle&=&|g_{R}\rangle_{1}|e_{0}\rangle_{2}|g_{L}\rangle_{3}|00\rangle,\nonumber\\
|\phi_{12}\rangle&=&|g_{R}\rangle_{1}|g_{R}\rangle_{2}|g_{L}\rangle_{3}|01\rangle,\nonumber\\
|\phi_{13}\rangle&=&|e_{0}\rangle_{1}|g_{R}\rangle_{2}|g_{L}\rangle_{3}|00\rangle,\nonumber\\
|\phi_{14}\rangle&=&|g_{0}\rangle_{1}|g_{R}\rangle_{2}|g_{L}\rangle_{3}|00\rangle,
\end{eqnarray}
The subscripts 1, 2, and 3 represent the atom 1, atom 2, and atom 3,
respectively. The state $|10\rangle$ ($|01\rangle$) denotes having
one left (right)-circularly photon while the state $|00\rangle$
describes none photon in the cavity. On the condition $\Omega\ll g$,
the Hilbert subspace is split into seven invariant Zeno subspaces
\cite{PF-PRL89,PF-JP196}
\begin{eqnarray}\label{3}
H_{p_{0}}=\{|\phi_{1}\rangle,
|\phi_{14}\rangle,|\varphi_{1}\rangle,|\varphi_{2}\rangle\},
H_{p_{1}}=\{|\varphi_{3}\rangle\},
H_{p_{2}}=\{|\varphi_{4}\rangle\}, H_{p_{3}}=\{|\varphi_{5}\rangle,
|\varphi_{6}\rangle\},\nonumber\\
H_{p_{4}}=\{|\varphi_{7}\rangle, |\varphi_{8}\rangle\},
H_{p_{5}}=\{|\varphi_{9}\rangle, |\varphi_{10}\rangle\},
H_{p_{6}}=\{|\varphi_{11}\rangle, |\varphi_{12}\rangle\},
\end{eqnarray}
where
\begin{eqnarray}\label{4}
|\varphi_{1}\rangle&=&\frac{1}{\sqrt{6}}(-|\phi_{3}\rangle+|\phi_{5}\rangle-|\phi_{7}\rangle+|\phi_{8}\rangle-|\phi_{10}\rangle+|\phi_{12}\rangle),\nonumber\\
|\varphi_{2}\rangle&=&\frac{1}{\sqrt{6}}(-|\phi_{2}\rangle+|\phi_{4}\rangle-|\phi_{6}\rangle+|\phi_{9}\rangle-|\phi_{11}\rangle+|\phi_{13}\rangle),\nonumber\\
|\varphi_{3}\rangle&=&\frac{1}{2\sqrt{3}}(|\phi_{2}\rangle+|\phi_{3}\rangle+|\phi_{4}\rangle+|\phi_{5}\rangle+|\phi_{6}\rangle{}
\nonumber\\
& & {}+|\phi_{7}\rangle+|\phi_{8}\rangle+|\phi_{9}\rangle+|\phi_{10}\rangle+|\phi_{11}\rangle+|\phi_{12}\rangle+|\phi_{13}\rangle),\nonumber\\
|\varphi_{4}\rangle&=&\frac{1}{2\sqrt{3}}(|\phi_{2}\rangle-|\phi_{3}\rangle+|\phi_{4}\rangle-|\phi_{5}\rangle+|\phi_{6}\rangle{}
\nonumber\\
& & {}-|\phi_{7}\rangle-|\phi_{8}\rangle+|\phi_{9}\rangle-|\phi_{10}\rangle+|\phi_{11}\rangle-|\phi_{12}\rangle+|\phi_{13}\rangle),\nonumber\\
|\varphi_{5}\rangle&=&\frac{1}{2\sqrt{2}}(|\phi_{2}\rangle-|\phi_{3}\rangle+|\phi_{5}\rangle-|\phi_{6}\rangle-|\phi_{9}\rangle+|\phi_{10}\rangle-|\phi_{12}\rangle+|\phi_{13}\rangle),\nonumber\\
|\varphi_{6}\rangle&=&\frac{1}{2\sqrt{6}}(|\phi_{2}\rangle+|\phi_{3}\rangle-2|\phi_{4}\rangle+|\phi_{5}\rangle+|\phi_{6}\rangle{}
\nonumber\\
& & {}-2|\phi_{7}\rangle-2|\phi_{8}\rangle+|\phi_{9}\rangle+|\phi_{10}\rangle-2|\phi_{11}\rangle+|\phi_{12}\rangle+|\phi_{13}\rangle),\nonumber\\
|\varphi_{7}\rangle&=&\frac{1}{2\sqrt{2}}(|\phi_{2}\rangle+|\phi_{3}\rangle-|\phi_{5}\rangle-|\phi_{6}\rangle-|\phi_{9}\rangle-|\phi_{10}\rangle+|\phi_{12}\rangle+|\phi_{13}\rangle),\nonumber\\
|\varphi_{8}\rangle&=&\frac{1}{2\sqrt{6}}(|\phi_{2}\rangle-|\phi_{3}\rangle-2|\phi_{4}\rangle-|\phi_{5}\rangle+|\phi_{6}\rangle{}
\nonumber\\
& & {}+2|\phi_{7}\rangle+2|\phi_{8}\rangle+|\phi_{9}\rangle-|\phi_{10}\rangle-2|\phi_{11}\rangle-|\phi_{12}\rangle+|\phi_{13}\rangle),\nonumber\\
|\varphi_{9}\rangle&=&\frac{1}{2\sqrt{12-6\sqrt{3}}}[-|\phi_{2}\rangle+|\phi_{3}\rangle+(1-\sqrt{3})|\phi_{4}\rangle-(\sqrt{3}-2)|\phi_{5}\rangle+(2-\sqrt{3})|\phi_{6}\rangle{}
\nonumber\\
& &
{}-(\sqrt{3}-1)|\phi_{7}\rangle+(\sqrt{3}-1)|\phi_{8}\rangle-(2-\sqrt{3})|\phi_{9}\rangle+(\sqrt{3}-2)|\phi_{10}\rangle{}
\nonumber\\
& & {}-(1-\sqrt{3})|\phi_{11}\rangle-|\phi_{12}\rangle+|\phi_{13}\rangle],\nonumber\\
|\varphi_{10}\rangle&=&\frac{1}{2\sqrt{12+6\sqrt{3}}}[-|\phi_{2}\rangle-|\phi_{3}\rangle+(1+\sqrt{3})|\phi_{4}\rangle-(\sqrt{3}+2)|\phi_{5}\rangle+(2+\sqrt{3})|\phi_{6}\rangle{}
\nonumber\\
& &
{}-(\sqrt{3}+1)|\phi_{7}\rangle+(\sqrt{3}+1)|\phi_{8}\rangle-(2+\sqrt{3})|\phi_{9}\rangle+(\sqrt{3}+2)|\phi_{10}\rangle{}
\nonumber\\
& & {}-(1+\sqrt{3})|\phi_{11}\rangle+|\phi_{12}\rangle+|\phi_{13}\rangle],\nonumber\\
|\varphi_{11}\rangle&=&\frac{1}{2\sqrt{12-6\sqrt{3}}}[-|\phi_{2}\rangle-|\phi_{3}\rangle+(1-\sqrt{3})|\phi_{4}\rangle+(\sqrt{3}-2)|\phi_{5}\rangle+(2-\sqrt{3})|\phi_{6}\rangle{}
\nonumber\\
& &
{}+(\sqrt{3}-1)|\phi_{7}\rangle-(\sqrt{3}-1)|\phi_{8}\rangle-(2-\sqrt{3})|\phi_{9}\rangle-(\sqrt{3}-2)|\phi_{10}\rangle{}
\nonumber\\
& & {}-(1-\sqrt{3})|\phi_{11}\rangle+|\phi_{12}\rangle+|\phi_{13}\rangle],\nonumber\\
|\varphi_{12}\rangle&=&\frac{1}{2\sqrt{12+6\sqrt{3}}}[-|\phi_{2}\rangle+|\phi_{3}\rangle+(1+\sqrt{3})|\phi_{4}\rangle+(\sqrt{3}+2)|\phi_{5}\rangle+(2+\sqrt{3})|\phi_{6}\rangle{}
\nonumber\\
& &
{}+(\sqrt{3}+1)|\phi_{7}\rangle-(\sqrt{3}+1)|\phi_{8}\rangle-(2+\sqrt{3})|\phi_{9}\rangle-(\sqrt{3}+2)|\phi_{10}\rangle{}
\nonumber\\
& &
{}-(1+\sqrt{3})|\phi_{11}\rangle-|\phi_{12}\rangle+|\phi_{13}\rangle],
\end{eqnarray}

corresponding to eigenvalues $\eta_{0}=0, \eta_{1}=2g,\eta_{2}=-2g,
\eta_{3}=-g, \eta_{4}=g, \eta_{5}=-\sqrt{3}g$ and
$\eta_{6}=\sqrt{3}g$ with the projections
\begin{eqnarray}\label{5}
P_{n}=\sum_{j}|\beta_{i,j}\rangle \langle \beta_{i,j}|,
(|\beta_{i,j}\rangle\in H_{P_{n}}).
\end{eqnarray}
Therefore the Hamiltonian of the current system is approximately
dominated by
\begin{eqnarray}\label{6}
 H_{total}&\cong& \sum_{n}(\eta_{n}P_{n}+P_{n}H_{laser}P_{n})
\cr\cr&=&2g|\varphi_{3}\rangle\langle
\varphi_{3}|-2g|\varphi_{4}\rangle\langle
\varphi_{4}|-g|\varphi_{5}\rangle\langle
\varphi_{5}|-g|\varphi_{6}\rangle\langle
\varphi_{6}|+g|\varphi_{7}\rangle\langle \varphi_{7}|{}
\nonumber\\
& & {}+g|\varphi_{8}\rangle\langle
\varphi_{8}|-\sqrt{3}g|\varphi_{9}\rangle\langle
\varphi_{9}|-\sqrt{3}g|\varphi_{10}\rangle\langle
\varphi_{10}|+\sqrt{3}g|\varphi_{11}\rangle\langle
\varphi_{11}|+\sqrt{3}g|\varphi_{12}\rangle\langle \varphi_{12}|{}
\nonumber\\
& & {}+\frac{1}{\sqrt{6}}(-\Omega|\phi_{1}\rangle\langle
\varphi_{2}|+\Omega|\phi_{14}\rangle\langle \varphi_{2}|+H.c.).
\end{eqnarray}

It is easily to find the Zeno subspace $H_{p_{0}}=
\{|\phi_{1}\rangle, |\phi_{14}\rangle, |\varphi_{1}\rangle,
|\varphi_{2}\rangle \}$ with eigenvalues $\eta_{0}=0$. Thus the
system state will evolve in that Zeno subspace with $H_{eff}$:
\begin{eqnarray}\label{7}
H_{eff}=\frac{1}{\sqrt{6}}(-\Omega|\phi_{1}\rangle\langle
\varphi_{2}|+\Omega|\phi_{14}\rangle\langle \varphi_{2}|+H.c.).
\end{eqnarray}
For an interaction time \emph{t}, the state of the whole system
becomes
\begin{eqnarray}\label{8}
|\Phi(t)\rangle=(\frac{1}{2}+\frac{1}{2}\cos{\frac{\sqrt{3}}{3}\Omega
t})|\phi_{1}\rangle-\frac{1}{2}(\cos{\frac{\sqrt{3}}{3}\Omega
t}-1)|\phi_{14}\rangle+i\frac{\sqrt{2}}{2}\sin{\frac{\sqrt{3}}{3}\Omega
t}|\varphi_{2}\rangle.
\end{eqnarray}
If we choose $\frac{\sqrt{3}}{3}\Omega t=\pi$ and the final state
becomes
$|\Phi(\frac{\sqrt{3}\pi}{\Omega})\rangle=|\phi_{14}\rangle$, one
will obtain the transform:
$|\phi_{1}\rangle=|g_{0}\rangle_{1}|g_{L}\rangle_{2}|g_{R}\rangle_{3}|00\rangle\rightarrow|\phi_{14}\rangle=|g_{0}\rangle_{1}|g_{R}\rangle_{2}|g_{L}\rangle_{3}|00\rangle$.

If the initial state is
$|g_{0}\rangle_{1}|g_{R}\rangle_{2}|g_{L}\rangle_{3}|00\rangle$, a
analogue method is utilized with the initial state
$|g_{0}\rangle_{1}|g_{L}\rangle_{2}|g_{R}\rangle_{3}|00\rangle$. As
a result, the interaction time is also $\frac{\sqrt{3}\pi}{\Omega}$
when the final state becomes
$|g_{0}\rangle_{1}|g_{L}\rangle_{2}|g_{R}\rangle_{3}|00\rangle$.

If the initial state is
$|g_{0}\rangle_{1}|g_{L}\rangle_{2}|g_{L}\rangle_{3}|00\rangle$, it
will evolve in a closed subspace spanned by
\begin{eqnarray}\label{9}
|\phi_{1}^{'}\rangle&=&|g_{0}\rangle_{1}|g_{L}\rangle_{2}|g_{L}\rangle_{3}|00\rangle,\nonumber\\
|\phi_{2}^{'}\rangle&=&|e_{0}\rangle_{1}|g_{L}\rangle_{2}|g_{L}\rangle_{3}|00\rangle,\nonumber\\
|\phi_{3}^{'}\rangle&=&|g_{L}\rangle_{1}|g_{L}\rangle_{2}|g_{L}\rangle_{3}|10\rangle,\nonumber\\
|\phi_{4}^{'}\rangle&=&|g_{L}\rangle_{1}|e_{0}\rangle_{2}|g_{L}\rangle_{3}|00\rangle,\nonumber\\
|\phi_{5}^{'}\rangle&=&|g_{L}\rangle_{1}|g_{R}\rangle_{2}|g_{L}\rangle_{3}|01\rangle,\nonumber\\
|\phi_{6}^{'}\rangle&=&|g_{L}\rangle_{1}|g_{L}\rangle_{2}|e_{0}\rangle_{3}|00\rangle,\nonumber\\
|\phi_{7}^{'}\rangle&=&|g_{L}\rangle_{1}|g_{L}\rangle_{2}|g_{R}\rangle_{3}|01\rangle,\nonumber\\
|\phi_{8}^{'}\rangle&=&|g_{R}\rangle_{1}|g_{L}\rangle_{2}|g_{L}\rangle_{3}|01\rangle.
\end{eqnarray}
On the condition $\Omega\ll g$, the Hilbert subspace is split into
five invariant Zeno subspaces \cite{PF-PRL89,PF-JP196}
\begin{eqnarray}\label{10}
H_{p_{0}}^{'}=\{|\phi_{1}^{'}\rangle, |\varphi_{1}^{'}\rangle\},
H_{p_{1}}^{'}=\{|\varphi_{2}^{'}\rangle, |\varphi_{3}^{'}\rangle\},
H_{p_{3}}^{'}=\{|\varphi_{4}^{'}\rangle, |\varphi_{5}^{'}\rangle\},
H_{p_{4}}^{'}=\{|\varphi_{6}^{'}\rangle\},
H_{p_{5}}^{'}=\{|\varphi_{7}^{'}\rangle\},
\end{eqnarray} where
\begin{eqnarray}\label{11}
|\varphi_{1}^{'}\rangle&=&\frac{1}{2}(-|\phi_{3}^{'}\rangle+|\phi_{5}^{'}\rangle+|\phi_{7}^{'}\rangle+|\phi_{8}^{'}\rangle),\nonumber\\
|\varphi_{2}^{'}\rangle&=&\frac{1}{2\sqrt{3}}(-|\phi_{2}^{'}\rangle+2|\phi_{4}^{'}\rangle-2|\phi_{5}^{'}\rangle-|\phi_{6}^{'}\rangle+|\phi_{7}^{'}\rangle+|\phi_{8}^{'}\rangle),\nonumber\\
|\varphi_{3}^{'}\rangle&=&\frac{1}{2}(-|\phi_{2}^{'}\rangle+|\phi_{6}^{'}\rangle-|\phi_{7}^{'}\rangle+|\phi_{8}^{'}\rangle),\nonumber\\
|\varphi_{4}^{'}\rangle&=&\frac{1}{2\sqrt{3}}(|\phi_{2}^{'}\rangle-2|\phi_{4}^{'}\rangle-2\phi_{5}^{'}\rangle+|\phi_{6}^{'}\rangle+|\phi_{7}^{'}\rangle+|\phi_{8}^{'}\rangle),\nonumber\\
|\varphi_{5}^{'}\rangle&=&\frac{1}{2}(|\phi_{2}^{'}\rangle-|\phi_{6}^{'}\rangle-|\phi_{7}^{'}\rangle+|\phi_{8}^{'}\rangle),\nonumber\\
|\varphi_{6}^{'}\rangle&=&\frac{1}{2\sqrt{6}}(-|\phi_{2}^{'}\rangle+3|\phi_{3}^{'}\rangle-2|\phi_{4}^{'}\rangle+|\phi_{5}^{'}\rangle-2|\phi_{6}^{'}\rangle+|\phi_{7}^{'}\rangle+|\phi_{8}^{'}\rangle),\nonumber\\
|\varphi_{7}^{'}\rangle&=&\frac{1}{2\sqrt{6}}(|\phi_{2}^{'}\rangle+3|\phi_{3}^{'}\rangle+2|\phi_{4}^{'}\rangle+|\phi_{5}^{'}\rangle+2|\phi_{6}^{'}\rangle+|\phi_{7}^{'}\rangle+|\phi_{8}^{'}\rangle),
\end{eqnarray}
corresponding to eigenvalues $\eta_{0}=0,
\eta_{1}^{'}=-g,\eta_{2}^{'}=g, \eta_{3}^{'}=-2g, \eta_{4}^{'}=2g,$
with the projections
\begin{eqnarray}\label{12}
P_{n}^{'}=\sum_{j}|\beta_{i,j}^{'}\rangle \langle \beta_{i,j}^{'}|,
(|\beta_{i,j}^{'}\rangle\in H_{P_{n}}^{'}).
\end{eqnarray}
Therefore the Hamiltonian of the current system is approximately
dominated by
\begin{eqnarray}\label{13}
 H_{total}^{'}&\cong& \sum_{n}(\eta_{n}^{'}P_{n}^{'}+P_{n}^{'}H_{laser}P_{n}^{'})
\cr\cr&=&-g|\varphi_{2}^{'}\rangle\langle
\varphi_{2}^{'}|-g|\varphi_{3}^{'}\rangle\langle
\varphi_{3}^{'}|+g|\varphi_{4}^{'}\rangle\langle
\varphi_{4}^{'}|+g|\varphi_{5}^{'}\rangle\langle
\varphi_{5}^{'}|-2g|\varphi_{6}^{'}\rangle\langle
\varphi_{6}^{'}|+2g|\varphi_{7}^{'}\rangle\langle \varphi_{7}^{'}|.
\end{eqnarray}

It is easily to find the Zeno subspace $H_{p_{0}}^{'}=
\{|\phi_{1}^{'}\rangle, |\varphi_{1}^{'}\rangle\}$ with eigenvalues
$\eta_{0}=0$. As $H_{total1}^{'}|\phi_{1}^{'}\rangle=0$, the final
state remains in the state
$|g_{0}\rangle_{1}|g_{L}\rangle_{2}|g_{L}\rangle_{3}|00\rangle$
without any change during the evolution.

If the initial state is
$|g_{0}\rangle_{1}|g_{R}\rangle_{2}|g_{R}\rangle_{3}|00\rangle$, a
analogue method is utilized with the initial state
$|g_{0}\rangle_{1}|g_{L}\rangle_{2}|g_{L}\rangle_{3}|00\rangle$. As
a result, the final state also remains in the state
$|g_{0}\rangle_{1}|g_{R}\rangle_{2}|g_{R}\rangle_{3}|00\rangle$
without any change.

Thus the three qubits can be described as follows:
\begin{eqnarray}\label{14}
|g_{R}\rangle_{1}|g_{R}\rangle_{2}|g_{R}\rangle_{3}|00\rangle&\rightarrow&|g_{R}\rangle_{1}|g_{R}\rangle_{2}|g_{R}\rangle_{3}|00\rangle,\nonumber\\
|g_{R}\rangle_{1}|g_{R}\rangle_{2}|g_{L}\rangle_{3}|00\rangle&\rightarrow&|g_{R}\rangle_{1}|g_{R}\rangle_{2}|g_{L}\rangle_{3}|00\rangle,\nonumber\\
|g_{R}\rangle_{1}|g_{L}\rangle_{2}|g_{R}\rangle_{3}|00\rangle&\rightarrow&|g_{R}\rangle_{1}|g_{L}\rangle_{2}|g_{R}\rangle_{3}|00\rangle,\nonumber\\
|g_{R}\rangle_{1}|g_{L}\rangle_{2}|g_{L}\rangle_{3}|00\rangle&\rightarrow&|g_{R}\rangle_{1}|g_{L}\rangle_{2}|g_{L}\rangle_{3}|00\rangle,\nonumber\\
|g_{0}\rangle_{1}|g_{R}\rangle_{2}|g_{R}\rangle_{3}|00\rangle&\rightarrow&|g_{0}\rangle_{1}|g_{R}\rangle_{2}|g_{R}\rangle_{3}|00\rangle,\nonumber\\
|g_{0}\rangle_{1}|g_{R}\rangle_{2}|g_{L}\rangle_{3}|00\rangle&\rightarrow&|g_{0}\rangle_{1}|g_{L}\rangle_{2}|g_{R}\rangle_{3}|00\rangle,\nonumber\\
|g_{0}\rangle_{1}|g_{L}\rangle_{2}|g_{R}\rangle_{3}|00\rangle&\rightarrow&|g_{0}\rangle_{1}|g_{R}\rangle_{2}|g_{L}\rangle_{3}|00\rangle,\nonumber\\
|g_{0}\rangle_{1}|g_{L}\rangle_{2}|g_{L}\rangle_{3}|00\rangle&\rightarrow&|g_{0}\rangle_{1}|g_{L}\rangle_{2}|g_{L}\rangle_{3}|00\rangle.
\end{eqnarray}
So we acquire the Fredkin gate from the Eq. (14) in the resonant
model.

\subsection{Under the large detuning condition}
As shown in Fig. 2, the transitions $|e_{0}\rangle_k \leftrightarrow
|g_{L}\rangle_k (|g_{R}\rangle_k)$ (\emph{k}=1,2,3) are coupled to
left-circularly (right-circularly) polarized cavity modes. The
transition $|e_{0}\rangle_1\leftrightarrow |g_{0}\rangle_1$ is
driven by classical laser field with Rabi frequency $\Omega$.
$\Delta$ denotes the detuning of the cavity modes and classical
laser field from the respective atomic transition. In the
interaction picture, the Hamiltonian for the whole system can be
written as ($\hbar = 1$)
\begin{eqnarray}\label{15}
H_{tot}&=&H_{ca}+H_{la}+H_{de},\nonumber\\
H_{ca}&=&\sum_{k=1}^{3}(g_{k,L}a_{L}|e_{0}\rangle_k\langle
g_{L}|+g_{k,R}a_{R}|e_{0}\rangle_k\langle g_{R}|)+H.c,\nonumber\\
H_{la}&=&\Omega(|e_{0}\rangle_{1}
\langle g_{0}|+|g_{0}\rangle_{1} \langle e_{0}|),\nonumber\\
H_{de}&=&\Delta(|e_{0}\rangle_{1}\langle
e_{0}|+|e_{0}\rangle_{2}\langle e_{0}|+|e_{0}\rangle_{3}\langle
e_{0}|).
\end{eqnarray}

The initial states including
$|g_{R}\rangle_{1}|g_{R}\rangle_{2}|g_{R}\rangle_{3}|00\rangle$,
$|g_{R}\rangle_{1}|g_{R}\rangle_{2}|g_{L}\rangle_{3}|00\rangle$,
$|g_{R}\rangle_{1}|g_{L}\rangle_{2}|g_{R}\rangle_{3}|00\rangle$, and
$|g_{R}\rangle_{1}|g_{L}\rangle_{2}|g_{L}\rangle_{3}|00\rangle$
remain unchanged during the time evolution because
$H_{tot}|g_{R}\rangle_{1}|g_{R}\rangle_{2}|g_{R}\rangle_{3}|00\rangle=0$,
$H_{tot}|g_{R}\rangle_{1}|g_{R}\rangle_{2}|g_{L}\rangle_{3}|00\rangle=0$,
$H_{tot}|g_{R}\rangle_{1}|g_{L}\rangle_{2}|g_{R}\rangle_{3}|00\rangle=0$,
and
$H_{tot}|g_{R}\rangle_{1}|g_{L}\rangle_{2}|g_{L}\rangle_{3}|00\rangle=0$.

If the initial state is
$|g_{0}\rangle_{1}|g_{L}\rangle_{2}|g_{R}\rangle_{3}|00\rangle$, the
whole system evolves in a closed subspace spanned by $\{
|\phi_{1}\rangle, |\phi_{2}\rangle, |\phi_{3}\rangle,
|\phi_{4}\rangle, |\phi_{5}\rangle, |\phi_{6}\rangle,
|\phi_{7}\rangle, |\phi_{8}\rangle, |\phi_{9}\rangle,
|\phi_{10}\rangle, |\phi_{11}\rangle, |\phi_{12}\rangle,
|\phi_{13}\rangle, |\phi_{14}\rangle \}$. Therefore we can rewrite
the above Hamiltonian in the ``$H_{ca}^{1}$'' representation:
\begin{eqnarray}\label{16}
H_{tot}^{1}&=&H_{ca}^{1}+H_{la}^{1}+H_{de}^{1},\nonumber\\
H_{ca}^{1}&=&2g|\varphi_{3}\rangle\langle
\varphi_{3}|-2g|\varphi_{4}\rangle\langle
\varphi_{4}|-g|\varphi_{5}\rangle\langle
\varphi_{5}|-g|\varphi_{6}\rangle\langle
\varphi_{6}|+g|\varphi_{7}\rangle\langle \varphi_{7}|{}
\nonumber\\
& & {}+g|\varphi_{8}\rangle\langle
\varphi_{8}|-\sqrt{3}g|\varphi_{9}\rangle\langle
\varphi_{9}|-\sqrt{3}g|\varphi_{10}\rangle\langle
\varphi_{10}|+\sqrt{3}g|\varphi_{11}\rangle\langle
\varphi_{11}|+\sqrt{3}g|\varphi_{12}\rangle\langle \varphi_{12}|,\nonumber\\
H_{la}^{1}&=&\frac{\Omega}{\sqrt{6}}|\varphi_{2}\rangle(\langle\phi_{14}|-\langle\phi_{1}|)+
\frac{\Omega}{2\sqrt{3}}(|\varphi_{3}\rangle+|\varphi_{4}\rangle)(\langle\phi_{14}|+\langle\phi_{1}|){}
\nonumber\\
& &
{}+\frac{\Omega}{2\sqrt{2}}(|\varphi_{5}\rangle+|\varphi_{7}\rangle)(\langle\phi_{14}|+\langle\phi_{1}|)+
\frac{\Omega}{2\sqrt{6}}(|\varphi_{6}\rangle+|\varphi_{8}\rangle)(\langle\phi_{14}|+\langle\phi_{1}|){}
\nonumber\\
& &
{}+\frac{\Omega}{2\sqrt{12-6\sqrt{3}}}(|\varphi_{9}\rangle+|\varphi_{11}\rangle)(\langle\phi_{14}|-\langle\phi_{1}|){}
\nonumber\\
& & {}+\frac{\Omega}{2\sqrt{12+6\sqrt{3}}}(|\varphi_{10}\rangle+|\varphi_{12}\rangle)(\langle\phi_{14}|-\langle\phi_{1}|)+H.c.,\nonumber\\
H_{de}^{1}&=&\Delta|\varphi_{2}\rangle\langle\varphi_{2}|+
\frac{\Delta}{2}(|\varphi_{3}\rangle+|\varphi_{4}\rangle)(\langle\varphi_{3}|+\langle\varphi_{4}|)+
\frac{\Delta}{2}(|\varphi_{5}\rangle+|\varphi_{7}\rangle)(\langle\varphi_{5}|+\langle\varphi_{7}|){}
\nonumber\\
& &
{}+\frac{\Delta}{2}(|\varphi_{6}\rangle+|\varphi_{8}\rangle)(\langle\varphi_{6}|+\langle\varphi_{8}|)+
\frac{\Delta}{2}(|\varphi_{9}\rangle+|\varphi_{11}\rangle)(\langle\varphi_{9}|+\langle\varphi_{11}|){}
\nonumber\\
& &
{}+\frac{\Delta}{2}(|\varphi_{10}\rangle+|\varphi_{12}\rangle)(\langle\varphi_{10}|+\langle\varphi_{12}|).
\end{eqnarray}
Furthermore, we assume that $U_{ca}^{1}=e^{-iH_{ca}^{1}t}$ is the
unitary time evolution operator with respect to the Hamiltonian
$H_{ca}^{1}$. After a calculation in the intermediate ``picture'',
we obtain
\begin{eqnarray}\label{17}
H_{la}^{1I}&=&U_{ca}^{1\dag}H_{la}^{1}U_{ca}^{1}
\cr\cr&=&\frac{\Omega}{\sqrt{6}}|\varphi_{2}\rangle(\langle\phi_{14}|-\langle\phi_{1}|)+
\frac{\Omega}{2\sqrt{3}}(e^{-i2gt}|\varphi_{3}\rangle+e^{i2gt}|\varphi_{4}\rangle)(\langle\phi_{14}|+\langle\phi_{1}|){}
\nonumber\\
& &
{}+\frac{\Omega}{2\sqrt{2}}(e^{igt}|\varphi_{5}\rangle+e^{-igt}|\varphi_{7}\rangle)(\langle\phi_{14}|+\langle\phi_{1}|)+
\frac{\Omega}{2\sqrt{6}}(e^{igt}|\varphi_{6}\rangle+e^{-igt}|\varphi_{8}\rangle)(\langle\phi_{14}|+\langle\phi_{1}|){}
\nonumber\\
& &
{}+\frac{\Omega}{2\sqrt{12-6\sqrt{3}}}(e^{i\sqrt{3}gt}|\varphi_{9}\rangle+e^{-i\sqrt{3}gt}|\varphi_{11}\rangle)(\langle\phi_{14}|-\langle\phi_{1}|){}
\nonumber\\
& & {}+
\frac{\Omega}{2\sqrt{12+6\sqrt{3}}}(e^{i\sqrt{3}gt}|\varphi_{10}\rangle+e^{-i\sqrt{3}gt}|\varphi_{12}\rangle)(\langle\phi_{14}|-\langle\phi_{1}|)+H.c.,\nonumber\\
H_{de}^{1I}&=&U_{ca}^{1\dag}H_{de}^{1}U_{ca}^{1}
\cr\cr&=&\Delta|\varphi_{2}\rangle\langle\varphi_{2}|+\frac{\Delta}{2}(|\varphi_{3}\rangle\langle\varphi_{3}|+|\varphi_{4}\rangle\langle\varphi_{4}|)
+\frac{\Delta}{2}(e^{-i4gt}|\varphi_{3}\rangle\langle\varphi_{4}|+e^{i4gt}|\varphi_{4}\rangle\langle\varphi_{3}|)
{}
\nonumber\\
& &
{}+\frac{\Delta}{2}(|\varphi_{5}\rangle\langle\varphi_{5}|+|\varphi_{7}\rangle\langle\varphi_{7}|)
+\frac{\Delta}{2}(e^{i2gt}|\varphi_{5}\rangle\langle\varphi_{7}|+e^{-i2gt}|\varphi_{7}\rangle\langle\varphi_{5}|)
{}
\nonumber\\
& &
{}+\frac{\Delta}{2}(|\varphi_{6}\rangle\langle\varphi_{6}|+|\varphi_{8}\rangle\langle\varphi_{8}|)
+\frac{\Delta}{2}(e^{i2gt}|\varphi_{6}\rangle\langle\varphi_{8}|+e^{-i2gt}|\varphi_{8}\rangle\langle\varphi_{6}|)
{}
\nonumber\\
& &
{}+\frac{\Delta}{2}(|\varphi_{9}\rangle\langle\varphi_{9}|+|\varphi_{11}\rangle\langle\varphi_{11}|)
+\frac{\Delta}{2}(e^{i2\sqrt{3}gt}|\varphi_{9}\rangle\langle\varphi_{11}|+e^{-i2\sqrt{3}gt}|\varphi_{11}\rangle\langle\varphi_{9}|)
{}
\nonumber\\
& &
{}+\frac{\Delta}{2}(|\varphi_{10}\rangle\langle\varphi_{10}|+|\varphi_{12}\rangle\langle\varphi_{12}|)
+\frac{\Delta}{2}(e^{i2\sqrt{3}gt}|\varphi_{10}\rangle\langle\varphi_{12}|+e^{-i2\sqrt{3}gt}|\varphi_{12}\rangle\langle\varphi_{10}|).
\end{eqnarray}
When the condition $\Omega\ll g$ is satisfied, we can safely discard
the terms in $H_{la}^{1I}$ and $H_{de}^{1I}$ with high oscillating
frequency $g$, $2g$, $4g$, $\sqrt{3}g$, and $2\sqrt{3}g$. Then we
move back to the original interaction picture from the intermediate
``picture'' and obtain
\begin{eqnarray}\label{18}
H^{1I}=\frac{\Omega}{\sqrt{6}}(|\phi_{14}\rangle-|\phi_{1}\rangle)\langle
\varphi_{2}|+\frac{\Omega}{\sqrt{6}}|\varphi_{2}\rangle(\langle
\phi_{14}|-|\langle \phi_{1}|)+\Delta|\varphi_{2}\rangle\langle
\varphi_{2}|.
\end{eqnarray}
One can find that the terms including
$|\varphi_{3}\rangle\langle\varphi_{3}|$,
$|\varphi_{4}\rangle\langle\varphi_{4}|$,
$|\varphi_{5}\rangle\langle\varphi_{5}|$,
$|\varphi_{6}\rangle\langle\varphi_{6}|$,
$|\varphi_{7}\rangle\langle\varphi_{7}|$,
$|\varphi_{8}\rangle\langle\varphi_{8}|$,
$|\varphi_{9}\rangle\langle\varphi_{9}|$,
$|\varphi_{10}\rangle\langle\varphi_{10}|$,
$|\varphi_{11}\rangle\langle\varphi_{11}|$, and
$|\varphi_{12}\rangle\langle\varphi_{12}|$ are discarded since they
are decoupled to the encoded qubit. If we consider
$|\Psi\rangle=\frac{1}{\sqrt{2}}(|\phi_{14}\rangle-|\phi_{1}\rangle)$
as a stable level, we can regard Eq. (17) as an effective
Hamiltonian of the two-level system. Thus the stable level is
coupled to the excited level $|\varphi_{2}\rangle$ with coupling
constant $\frac{\Omega}{\sqrt{3}}$ and detuning $\Delta$. On the
large detuning condition ($\Delta\gg\Omega$), there is no transition
between $|\Psi\rangle$ and $|\varphi_{2}\rangle$, only the stark
shift contributes to the variation of energy for each level
\cite{DF-CJP85}. By adiabatically eliminating the excited state
$|\varphi_{2}\rangle$, we obtain the final effective Hamiltonian
\begin{eqnarray}\label{19}
H_{eff}^{1I}=-\frac{\Omega^{2}}{6\Delta}(|\phi_{1}\rangle\langle\phi_{1}|+|\phi_{14}\rangle\langle\phi_{14}|)+
\frac{\Omega^{2}}{6\Delta}(|\phi_{1}\rangle\langle\phi_{14}|+|\phi_{1}\rangle\langle\phi_{14}|).
\end{eqnarray}
For an interaction time $t$, the state of the whole system becomes
\begin{eqnarray}\label{20}
|\Phi(t)\rangle=\frac{1}{2}(1+\cos{\frac{\Omega^{2}t}{3\Delta}}+i\sin{\frac{\Omega^{2}t}{3\Delta}})|\phi_{1}\rangle+
\frac{1}{2}(1-\cos{\frac{\Omega^{2}t}{3\Delta}}-i\sin{\frac{\Omega^{2}t}{3\Delta}})|\phi_{14}\rangle.
\end{eqnarray}
If we choose $\frac{\Omega^{2}t}{3\Delta}=\pi$ and the final state
becomes
$|\Phi(\frac{3\Delta\pi}{\Omega^{2}})\rangle=|\phi_{14}\rangle$, one
will obtain the transform:
$|\phi_{1}\rangle=|g_{0}\rangle_{1}|g_{L}\rangle_{2}|g_{R}\rangle_{3}|00\rangle\rightarrow|\phi_{14}\rangle=|g_{0}\rangle_{1}|g_{R}\rangle_{2}|g_{L}\rangle_{3}|00\rangle$.

If the initial state is
$|g_{0}\rangle_{1}|g_{R}\rangle_{2}|g_{L}\rangle_{3}|00\rangle$, a
analogue method is utilized with the initial state
$|g_{0}\rangle_{1}|g_{L}\rangle_{2}|g_{R}\rangle_{3}|00\rangle$. As
a result, the interaction time is also
$t=\frac{3\Delta\pi}{\Omega^{2}}$ when the final state becomes
$|g_{0}\rangle_{1}|g_{L}\rangle_{2}|g_{R}\rangle_{3}|00\rangle$.

If the initial state is
$|g_{0}\rangle_{1}|g_{L}\rangle_{2}|g_{L}\rangle_{3}|00\rangle$, the
whole system evolves in a closed subspace spanned by $
\{|\phi_{1}^{'}\rangle, |\phi_{2}^{'}\rangle, |\phi_{3}^{'}\rangle,
|\phi_{4}^{'}\rangle, |\phi_{5}^{'}\rangle, |\phi_{6}^{'}\rangle,
|\phi_{7}^{'}\rangle, |\phi_{8}^{'}\rangle\}$. Then we can rewrite
the above Hamiltonian in the ``$H_{ca}^{2}$'' representation:
\begin{eqnarray}\label{21}
H_{tot}^{2}&=&H_{ca}^{2}+H_{la}^{2}+H_{de}^{2},\nonumber\\
H_{ca}^{2}&=&-g|\varphi_{2}^{'}\rangle\langle
\varphi_{2}^{'}|-g|\varphi_{3}^{'}\rangle\langle
\varphi_{3}^{'}|+g|\varphi_{4}^{'}\rangle\langle
\varphi_{4}^{'}|+g|\varphi_{5}^{'}\rangle\langle
\varphi_{5}^{'}|-2g|\varphi_{6}^{'}\rangle\langle
\varphi_{6}^{'}|+2g|\varphi_{7}^{'}\rangle\langle \varphi_{7}^{'}|,\nonumber\\
H_{la}^{2}&=&\frac{\Omega}{2\sqrt{3}}(-|\varphi_{2}^{'}\rangle+|\varphi_{4}^{'}\rangle)\langle\phi_{1}^{'}|
+\frac{\Omega}{2}(-|\varphi_{3}^{'}\rangle+|\varphi_{5}^{'}\rangle)\langle\phi_{1}^{'}|
+\frac{\Omega}{2\sqrt{6}}(-|\varphi_{6}^{'}\rangle+|\varphi_{7}^{'}\rangle)\langle\phi_{1}^{'}|,\nonumber\\
H_{de}^{2}&=&\frac{\Delta}{2}(|\varphi_{2}^{'}\rangle-|\varphi_{4}^{'}\rangle)(\langle
\varphi_{2}^{'}|-\langle
\varphi_{4}^{'}|)+\frac{\Delta}{2}(|\varphi_{3}^{'}\rangle-|\varphi_{5}^{'}\rangle)(\langle
\varphi_{3}^{'}|-\langle \varphi_{5}^{'}|){}
\nonumber\\
& &
{}+\frac{\Delta}{2}(|\varphi_{6}^{'}\rangle-|\varphi_{7}^{'}\rangle)(\langle
\varphi_{6}^{'}|-\langle \varphi_{7}^{'}|).
\end{eqnarray}
A analogue method is utilized with the Eq. (16) - Eq. (18). We can
obtain that the final effective Hamiltonian $H_{eff}^{2I}$ has no
effect on the time evolution of initial state
$|g_{0}\rangle_{1}|g_{L}\rangle_{2}|g_{L}\rangle_{3}|00\rangle$.
Namely,
\begin{eqnarray}\label{22}
H_{eff}^{2I}|g_{0}\rangle_{1}|g_{L}\rangle_{2}|g_{L}\rangle_{3}|00\rangle=0.
\end{eqnarray}
Therefore, the initial state
$|g_{0}\rangle_{1}|g_{L}\rangle_{2}|g_{L}\rangle_{3}|00\rangle$
remains unchanged during the time evolution.

Specially, it is a similar case to the initial state
$|g_{0}\rangle_{1}|g_{R}\rangle_{2}|g_{R}\rangle_{3}|00\rangle$. The
final state does not have any change during the evolution. As a
result, we can also acquire the Fredkin gate in the detuning model.

\section{numerical analysis and discussions}

All the above derivations are based on the ideal case that the
influence of decoherence induced by cavity decay and atomic
spontaneous emission on the time evolution of system is omitted. We
here utilize a quantum jump approach
\cite{MB-RMP101,GH-QSO8,JY-PRL68} to discuss the influence of
decoherence on the time evolution of system. If no photon is
detected through the leakage from the cavity and the atomic
spontaneous emission, the time evolution of system is dominated by
the conditional Hamiltonian
\begin{eqnarray}\label{24}
H_{cond}=H_{total}-\sum_{k=1}^{3}\sum_{j=L,R}(i\frac{\gamma_{j}}{2}|e_{0}\rangle_{k}\langle
e_{0}|+i\frac{\kappa_{j}}{2}a_{j}^{\dag}a_{j}),
\end{eqnarray}
where $\gamma_{j}$ is spontaneous emission rate for the excited
state $|e_{0}\rangle$ and $\kappa_{j}$ is decay rate of the
corresponding cavity mode. Without loss of generality, we set
$\gamma_{j}=\gamma$ and $\kappa_{j}=\kappa$. Assuming the initially
state
$|\Phi(0)\rangle=\frac{1}{2\sqrt{2}}(g_{0}+g_{R})\otimes(g_{L}+g_{R})\otimes(g_{L}+g_{R})$,
the time evolution of system is
\begin{eqnarray}\label{25}
|\Phi(t)\rangle=\frac{U_{cond}(t)|\Phi(0)\rangle}{\sqrt{P_{suc}(t)}},
\end{eqnarray}
where $U_{cond}(t)=e^{-iH_{cond}t}$ is the time evolution operator
for $H_{cond}$,
$P_{suc}(t)=\langle\Phi(0)|U_{cond}^{\dag}(t)U_{cond}(t)|\Phi(0)\rangle$
represents that no photon has been emitted at time \emph{t}. The
computation will fail and has to be repeated if photons are emitted.
To some extent, it can be made up by monitoring photon emissions
with good detectors.

Fig. 3 (Fig. 4) shows the relation among the fidelity $F$ (the
success probability $P$) of the Fredkin gate and cavity decay and
atomic spontaneous emission with the other parameter chosen as
$\Omega=0.03g$ in the resonant model. One can see from Fig. 3 that
with the increasing of cavity decay and atomic spontaneous emission,
the fidelity $F$ of the Fredkin gate will decrease. If we set
$\gamma/g = 0.1$ and $\kappa/g = 0.1$, one can obtain the value of
fidelity $F = 77.78\%$ and success probability $P = 88.66\%$.
However, we can see from Fig. 3 that this model is robust against
cavity decay due to the fidelity $F$ is about 99.75\% even though
cavity decay $\kappa/g = 0.1$ when spontaneous emission $\gamma/g =
0$. That is due to the fact that the system evolves in a closed
subspace where the cavity mode field is not excitation. Thus the
main decoherence of the resonant model is spontaneous emission
because the evolution of whole system involves excited states of the
atoms. As a result, the resonant model is suitable for small atomic
spontaneous emission rate and large cavity decay rate.

Fig. 5 (Fig. 6) shows the relation among the fidelity $F$ (the
success probability $P$) of the Fredkin gate and cavity decay and
atomic spontaneous emission with the other parameters chosen as
$\Omega=0.03g$ and $\Delta=0.3g$ in the large detuning model. One
can see from Fig. 5 (Fig. 6) that with the increasing of cavity
decay and atomic spontaneous emission, the fidelity $F$ (the success
probability $P$) of the Fredkin gate will decrease. If we set
$\gamma/g = 0.1$ and $\kappa/g = 0.1$ ($g^{2}=100\kappa\gamma$), one
can obtain the value of fidelity $F = 97.83\%$ and success
probability $P = 84.60\%$. As a result, the fidelity $F$ is
insensitive to cavity decay. That is also due to the fact that the
system evolves in a closed subspace where the cavity mode field is
not excitation. On the other hand, the fidelity $F$ is also
insensitive to atomic spontaneous emission because of the presence
of the large detuning, eliminating the excited state of atoms
adiabatically. As a result, the lage detuning model is suitable for
moderate values of atomic spontaneous emission rate and cavity decay
rate and the lifetime of the states should be long enough because
the interaction time needs longer compared to the resonant model.
Thus high fidelity of the proposed Fredkin gate can be obtained for
the optimal value of the large detuning and also depends on the
practical experimental parameters restriction of cavity decay and
atomic spontaneous emission.

\section{EXPERIMENTAL FEASIBILITY AND CONCLUSIONS}

Finally, we give a brief discussion on the experimental feasibility
of both models. The atomic configuration involved in our models can
be implemented with $^{87}\textrm{Rb}$. The relevant atomic levels
are shown in Fig. 7. Each atom is assumed to be coupled to
$\sigma^{+}$ and $\sigma^{-}$ -polarized photon modes of the
bi-modal cavity and the first atom is coupled to an external
$\pi$-polarized classical field individually. In our models, all the
encoded qubit states are low-energy states and the cavity modes are
almost in the vacuum state during the time evolution. On the other
hand, it is feasible with the parameters $g =2\pi\times750$ MHz,
$\gamma = 2\pi\times2.62$ MHz, $\kappa = 2\pi\times3.5$ MHz in an
optical cavity with the wavelength in the region 630 - 850 nm in
recent experiments \cite{SM-PRA71,JR-PRA67}. In the resonant model,
the fidelity will be about 99.73\% and the interaction time
$t\simeq0.0385\mu$s. We will obtain a high fidelity about 99.93\%
and the interaction time $t\simeq0.667\mu$s in the detuning model.
Therefore, it allows construction of an atomic system for quantum
computation in the presence of decoherence.

In summary, we have studied one-step implementation of the Fredkin
gate in a bi-modal cavity under both resonant and large detuning
conditions based on quantum Zeno dynamics, which reduces the
complexity of experiment operations. The principle implementing the
Fredkin gate is quite different in both models which each of them
has its respective advantages. Therefore, it provides a flexibility
to adapt the proposed model to different experimental apparatus. In
the resonant model, it is robust against the cavity decay because
the state keeps in a subspace without exciting the cavity field
during the evolution and the interaction time needs rather short
because of the resonant interaction. However atomic spontaneous
emission is the main decoherence in this model. In the large
detuning model, it is insensitive to the influence of decoherence
caused by decay of the cavity modes and spontaneous of the excited
states due to the large detuning condition. However the time is much
longer than the time required in the resonant model on the same Rabi
frequency $\Omega$ and the same coupling strength $g$. In addition,
we have also briefly discussed the influence of decoherence induced
by cavity decay and atomic spontaneous emission by numerical
calculation. Therefore, we hope that with the current experimental
technology it may be possible to implement the Fredkin gate in this
paper.

\acknowledgements

This work was supported by the Natural Science Foundation of Fuzhou
University of China under grant no 022264 and no 2010-XQ-28, the
funds from Education Department of Fujian Province of China under
grant no JB08010, no JA10009 and no JA10039, the National Natural
Science Foundation of Fujian Province of China under grant No.
2009J06002 and no 2010J01006, the National Natural Science
Foundation of China under grant no 11047122, no 10875020 and
10974028, Doctoral Foundation of the Ministry of Education of China
under grant no 20093514110009, and China Postdoctoral Science
Foundation under grant no 20100471450.

\newpage

FIG. 1. The level configuration of the atoms (the resonant model).

FIG. 2. The level configuration of the atoms (the large detuning
model).

FIG. 3. The influence of cavity decay and atomic spontaneous
emission on the fidelity of the Fredkin gate (the resonant model).
The other parameter is $\Omega=0.03g$.

FIG. 4. The influence of cavity decay and atomic spontaneous
emission on the success probability of the Fredkin gate (the
resonant model). The other parameter is $\Omega=0.03g$.

FIG. 5. The influence of cavity decay and atomic spontaneous
emission on the fidelity of the Fredkin gate (the large detuning
model). The other parameters are $\Omega=0.03g, \Delta=0.3g$.

FIG. 6. The influence of cavity decay and atomic spontaneous
emission on the success probability of the Fredkin gate (the large
detuning model). The other parameters are $\Omega=0.03g,
\Delta=0.3g$.

FIG. 7. The energy levels of $^{87}\textrm{Rb}$ (left side denotes
the resonant model and right side denotes the large detuning model).

\newpage

\begin {figure}
\scalebox{0.8}{\includegraphics {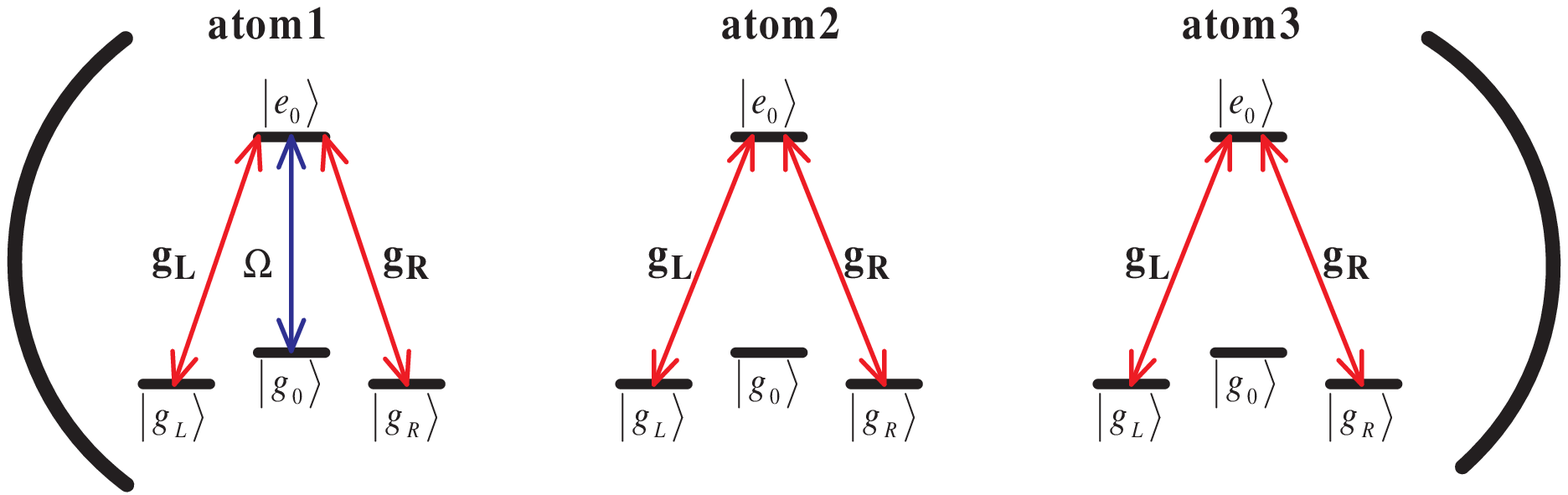}} \caption{} \label {Fig1}
\end{figure}

\begin {figure}
\scalebox{0.8}{\includegraphics {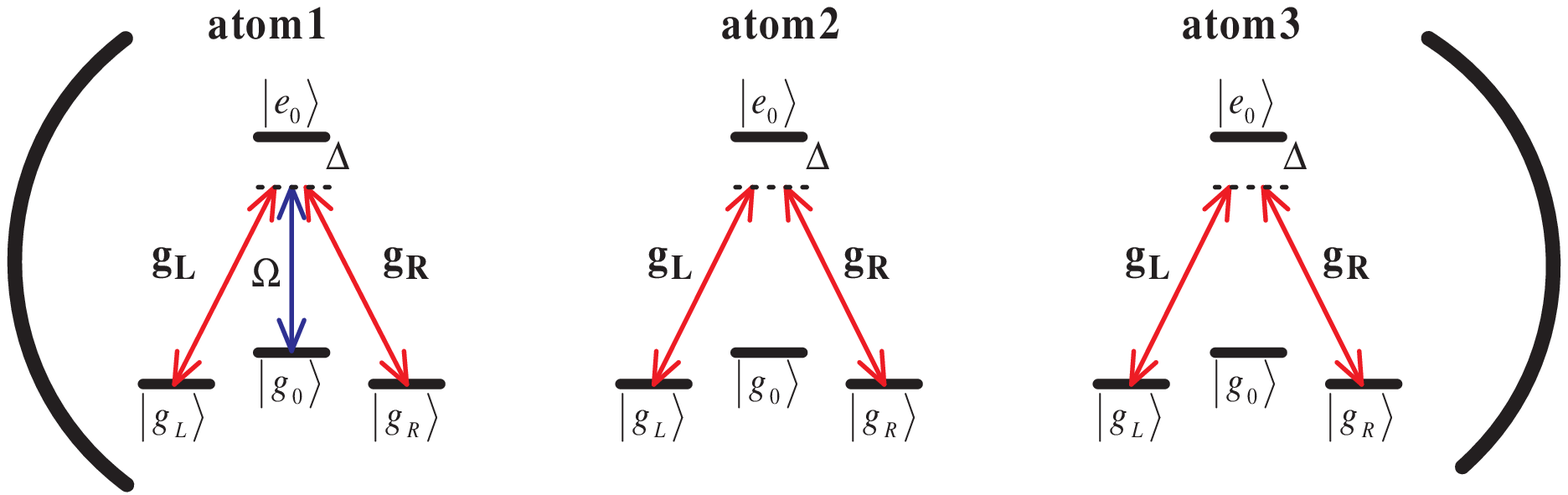}} \caption{} \label {Fig2}
\end{figure}

\begin {figure}
\scalebox{0.8}{\includegraphics {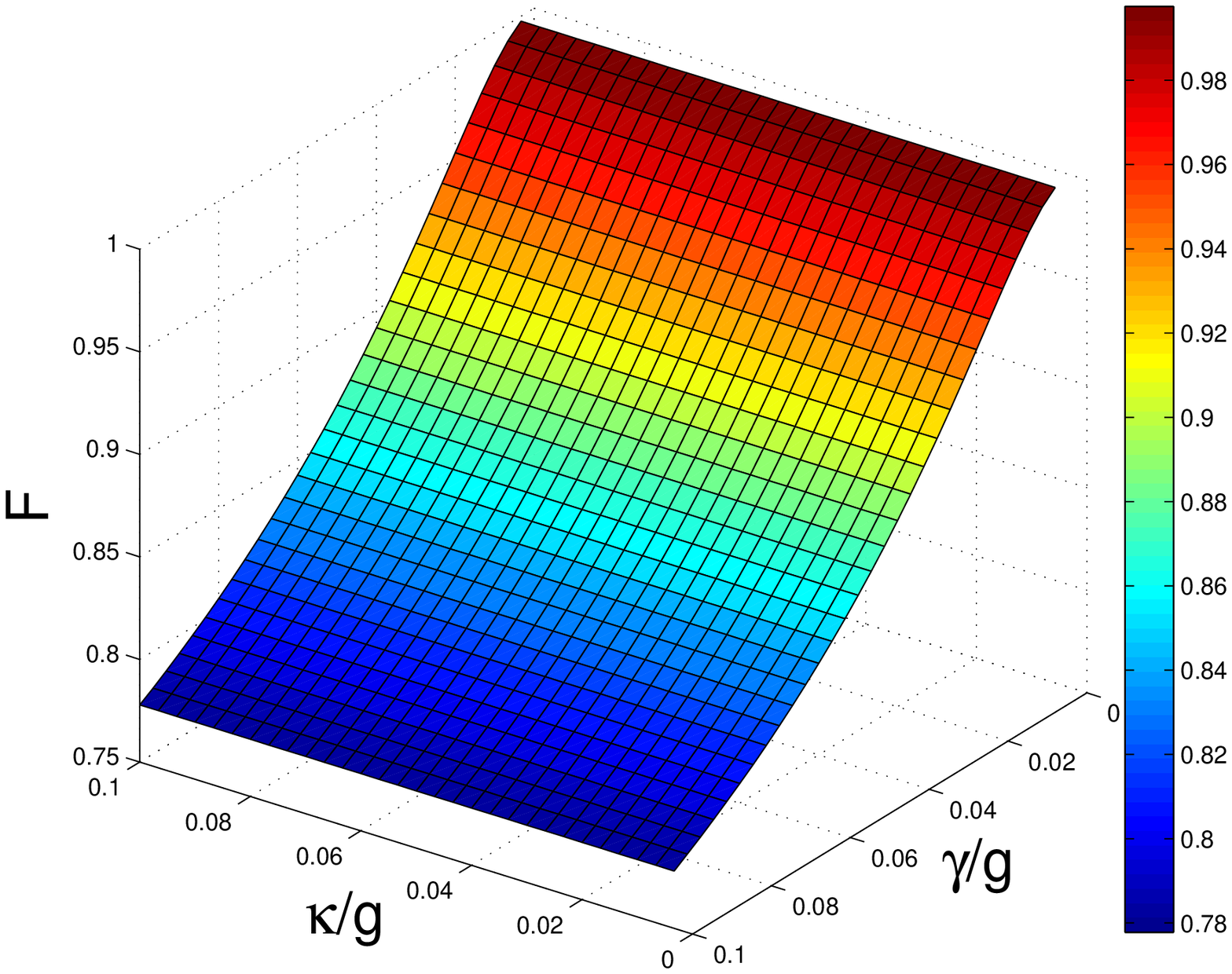}} \caption{} \label {Fig3}
\end{figure}

\begin {figure}
\scalebox{0.8}{\includegraphics {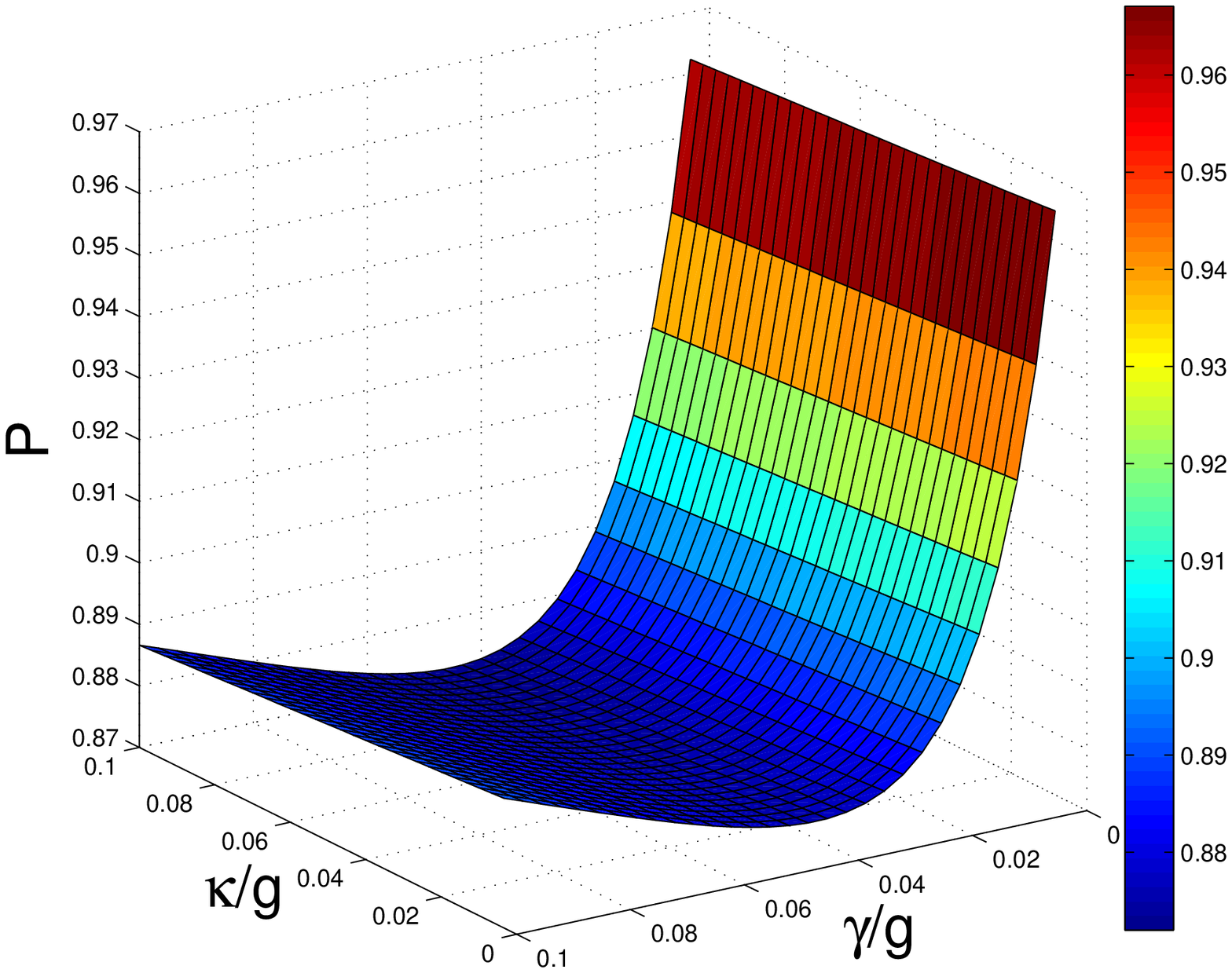}} \caption{} \label {Fig4}
\end{figure}

\begin {figure}
\scalebox{0.8}{\includegraphics {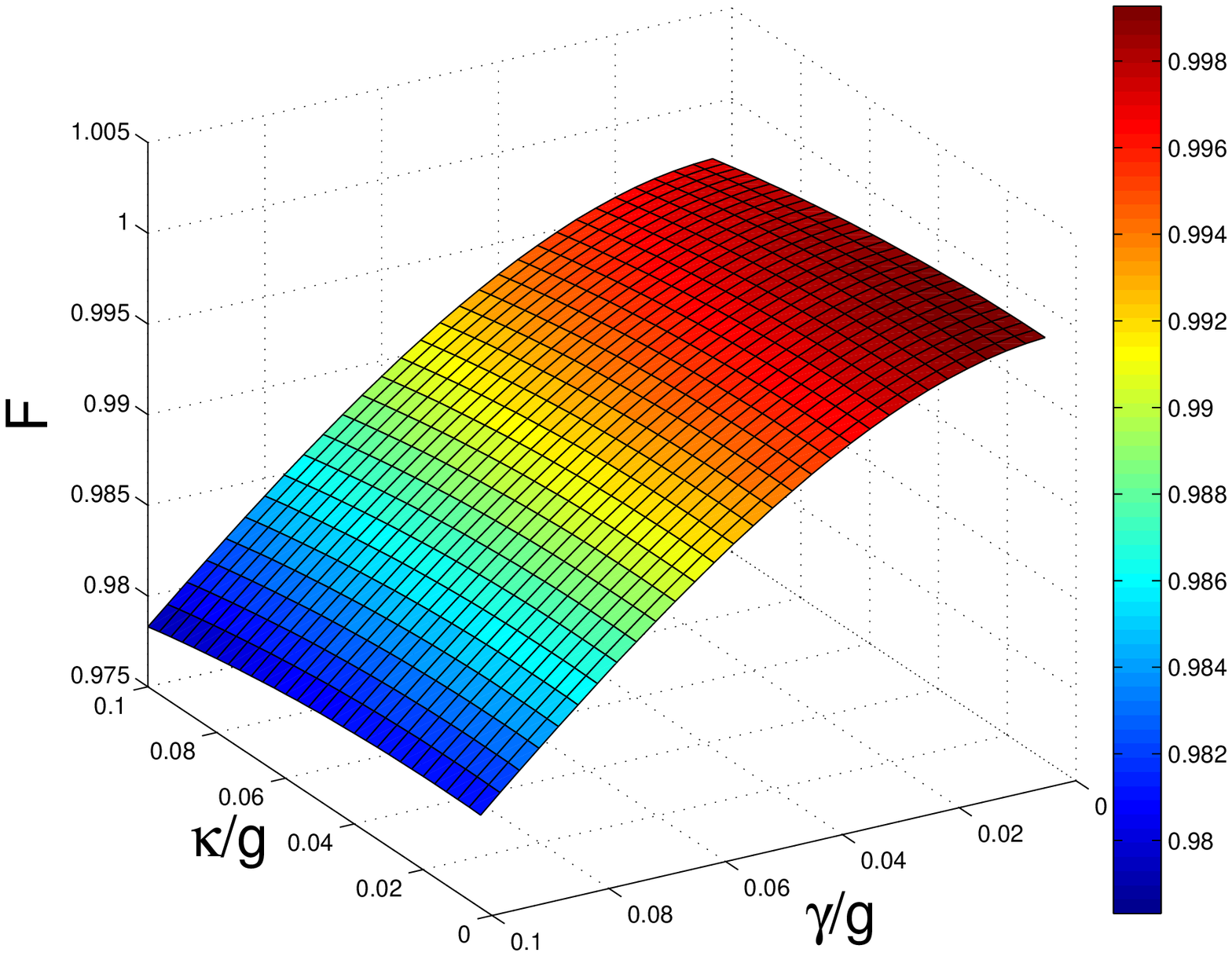}} \caption{} \label {Fig5}
\end{figure}

\begin {figure}
\scalebox{0.8}{\includegraphics {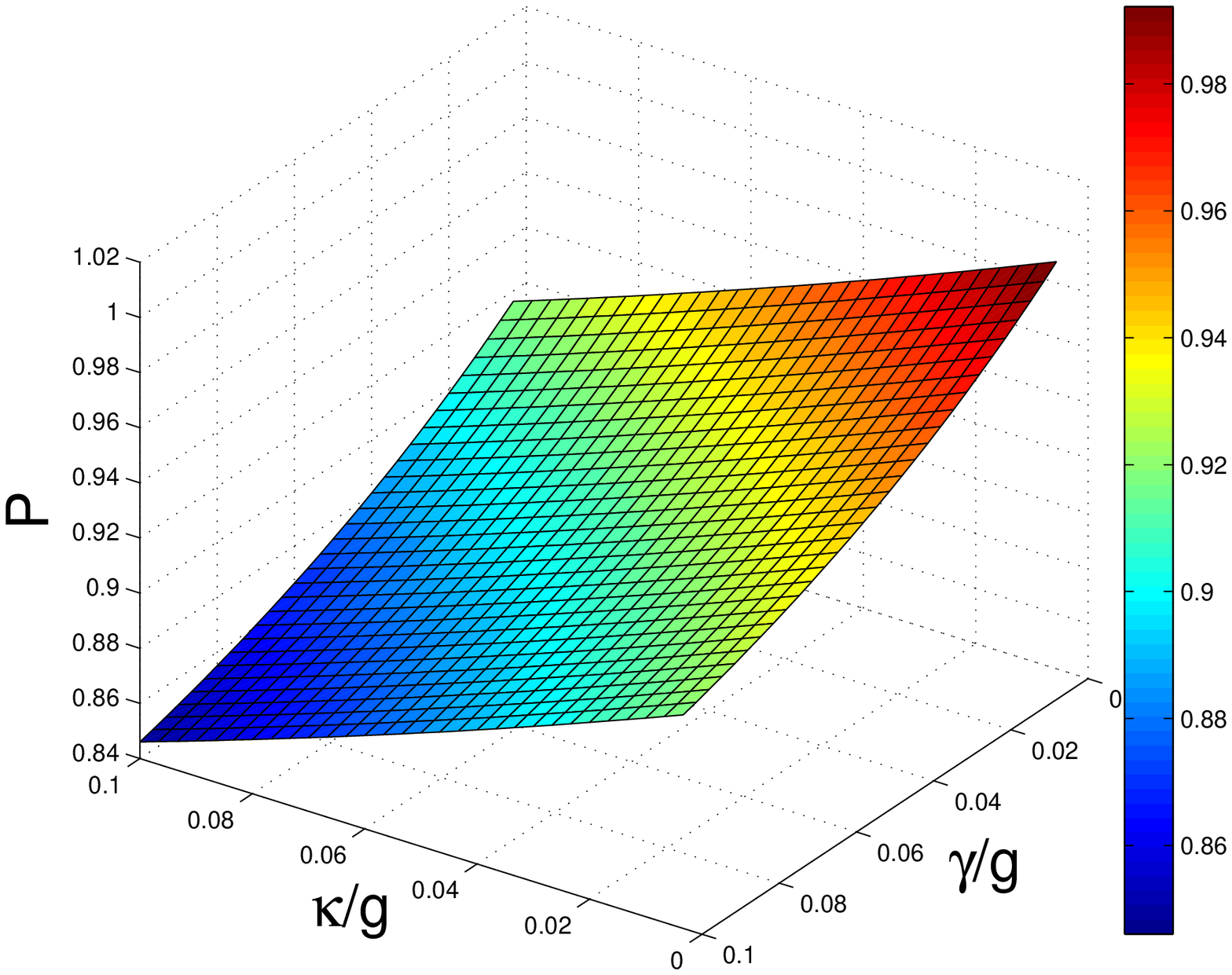}} \caption{} \label {Fig6}
\end{figure}

\begin {figure}
\scalebox{0.8}{\includegraphics {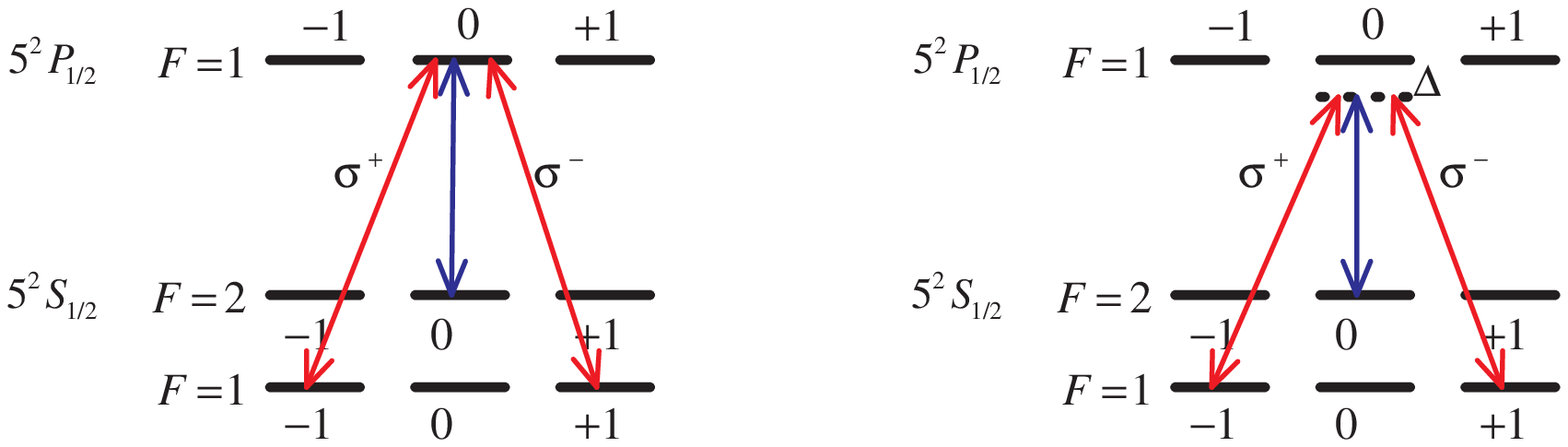}} \caption{} \label {Fig7}
\end{figure}

 \end{document}